\documentclass[12pt]{article}
\usepackage{graphicx}
\usepackage{calrsfs}
\usepackage{geometry}                		% See geometry.pdf to learn the layout options. There are lots.
\usepackage{graphicx}				% Use pdf, png, jpg, or eps§ with pdflatex; use eps in DVI mode
\usepackage{amssymb}
\usepackage[pdfstartview=FitH,
colorlinks=true,linkcolor=red,anchorcolor=black,citecolor=blue,urlcolor=black,filecolor=blue]{hyperref}
\usepackage{amstext}
\usepackage{caption}
\usepackage{etoolbox}
\usepackage{makeidx}
\include{epsf}
\usepackage[toc,page]{appendix}
\usepackage{amsfonts}
\usepackage{authblk} 
\usepackage{tcolorbox}
\usepackage{sectsty}
\usepackage{amsmath,amssymb,epsfig}
\usepackage{amscd}
\usepackage{amsthm}
\usepackage{braket}
\usepackage{amsmath}
\usepackage{xcolor}
\usepackage{amssymb}
\usepackage{mathrsfs}
\usepackage{setspace}
\usepackage{url}
\usepackage{dsfont}
\usepackage[applemac]{inputenc}
\usepackage[english]{babel}
\usepackage{enumitem} 
\usepackage{verbatim}
\usepackage{slashed}
\usepackage[]{latexsym}
\usepackage{mathtools}
\usepackage{subcaption}
\usepackage{ mathrsfs }
\usepackage{braket}
\usepackage{caption}
\usepackage{cite}
\usepackage{dsfont}
\usepackage{mathpazo}
\usepackage{hyperref}
\hypersetup{
pdftitle={},%
pdfauthor={},%
pdfsubject={},%
pdfkeywords={},%
colorlinks=true,%
linkcolor=blue,%
citecolor=red,%
linktocpage=true,%
pageanchor=true
}

\newcommand{\be}{\begin{equation}}
\newcommand{\ee}{\end{equation}}
\def\beqa{\begin{eqnarray}}
\def\eeqa{\end{eqnarray}}
\def\bean{\begin{eqnarray*}}
\def\eean{\end{eqnarray*}}

\newcommand{\R}{\mathbb{R}}
\newcommand{\C}{\mathbb{C}}

\newcommand{\largewedge}{\mbox{\Large $\wedge$}}

\newcommand{\dd}{{\mathrm{d}}}
\theoremstyle{definition}
\newtheorem{definition}{Definition}[section]
\newtheorem{theorem}{Theorem}[section]

\newtheorem{example}{Example}[section]
\newtheorem{corollary}{Corollary}[section]

\newcommand{\eqn}[1]{(\ref{#1})}
\newcommand{\del}{\partial}

\renewenvironment{thebibliography}[1]
         {\section*{References}\frenchspacing\small
          \begin{list}{[\arabic{enumi}]}
         {\usecounter{enumi}\parsep=2pt\topsep 0pt
         \settowidth{\labelwidth}{[#1]}
         \leftmargin=\labelwidth\advance\leftmargin\labelsep
         \rightmargin=0pt\itemsep=1pt\sloppy}}{\end{list}}

 \numberwithin{equation}{section}

\title{\textbf{Jacobi Sigma Models and Twisted Jacobi Structures}\vspace{0.5cm}}
\date{}

\author[1]{Francesco Bascone}
\author[1]{Franco Pezzella}
\author[1,2]{Patrizia Vitale}
\affil[ ]{}
\affil[1]{\textit{\footnotesize INFN-Sezione di Napoli, Complesso Universitario di Monte S. Angelo Edificio 6, via Cintia, 80126 Napoli, Italy.}}
\affil[2]{\textit{\footnotesize Dipartimento di Fisica ``E. Pancini'', Universit\`a di Napoli Federico II, Complesso Universitario di Monte S. Angelo Edificio 6, via Cintia, 80126 Napoli, Italy.}}
\affil[ ]{}
\affil[ ]{\footnotesize e-mail: \texttt{francesco.bascone@na.infn.it, franco.pezzella@na.infn.it, patrizia.vitale@na.infn.it}}
\begin{document}
\maketitle
\begin{abstract}
\small 
Jacobi sigma models are two-dimensional topological non-linear  field theories which are associated with Jacobi structures. The latter can be considered as a generalization of Poisson structures. After reviewing the main properties and peculiarities of these models, we focus on the twisted version in which a Wess-Zumino term is included. This modification allows for the target space to be a twisted Jacobi manifold. We discuss in  particular  the model on the sphere $S^5$.
%, which is relevant in the context of $AdS_5 \times S^5$ superstring theory, 
%but here considered 
%as an example of twisted contact manifold.

\end{abstract}

\newpage
\tableofcontents

\section{Introduction}\label{secintroduction}

Jacobi sigma models  \cite{Bascone:2020drt, Chatzistavrakidis:2020gpv} are topological non-linear gauge field theories generalizing the well-known Poisson sigma model (PSM). They derive their name from the properties of the target space. This is  a    Jacobi manifold, namely a  differential manifold endowed with a Jacobi bracket, introduced by  Kirillov \cite{KirillovJacobi} as the most general Lie bracket one may define on the algebra of functions when an additional locality requirement is assumed.  

As a generalization of   Poisson tensors which satisfy the condition   $[\Pi, \Pi]_S=0$, the latter being the Schouten bracket,   Jacobi structures are specified by a bivector field $\Lambda$ and a vector field $E$, satisfying 
\begin{equation}
[\Lambda, \Lambda]_S=2 E \wedge \Lambda, \quad [E, \Lambda]_S=0.
\end{equation}
A manifold equipped with such a structure is  a Jacobi manifold. Clearly,  Poisson manifolds are a special case with  $E=0$. The  Jacobi bracket is then defined as
\begin{equation}
\{f,g \}_{J}=\Lambda(df,dg)+\iota_E(fdg-gdf),
\end{equation}
for any smooth functions $f$, $g$ on $M$ and $\iota_E$ the contraction operator. This bracket satisfies the Jacobi identity but violates the Leibniz rule, so unlike the Poisson bracket it does not represent a derivation of the pointwise product among functions. More details  on Jacobi manifolds shall be given in Sec. \ref{Jacobiman}.

The Jacobi sigma model (JSM) proposed in \cite{Bascone:2020drt, Chatzistavrakidis:2020gpv} generalizes the construction of the Poisson sigma model by introducing another field which is necessary in order to   take into account    the new background field $E$. The model is defined on a two-dimensional manifold, the worldsheet $\Sigma$,  %(the worldsheet in the  language of strings), 
with the  target  $M$ being a Jacobi manifold. The field variables are represented by the triple $(X, \eta, \lambda)$, with $X: \Sigma \to M$ the usual embedding map, while  $(\eta, \lambda)$ are elements of $\Omega^1\left(\Sigma, X^*\left(T^* M \oplus \mathbb{R}\right)\right)$,  namely one forms on $\Sigma$ which take value in the pullback of $T^* M \oplus \mathbb{R}$. The latter is the 1-jet bundle of real functions on $M$, usually indicated with $J^1M$.

The action in ref. \cite{Bascone:2020drt} is given by
\be\label{jacobac}
S[X,(\eta, \lambda)]= \int_{\Sigma}\left[\eta_i\wedge dX^i+\frac{1}{2}\Lambda^{ij}(X) \eta_i\wedge\eta_j -E^i(X)\eta_i\wedge\lambda\right]
\ee
which reduces to the PSM for $E=0$. The proposal of \cite{Chatzistavrakidis:2020gpv} is slightly different because it contains a scalar field coming from a dimensional reduction in  target space. 
The canonical formulation  has been studied in  detail in  \cite{Bascone:2021njt}, showing that the model   exhibits first and second class constraints, the former being the generators of gauge transformations; the reduced phase space 
% which is obtained as the constrained manifold modulo gauge symmetries, 
has finite dimension $2\text{dim}M-2$.  It is also shown that it is possible to establish a homomorphism between the algebra of gauge transformations and the algebra of sections of   $J^1M$. This generalises an analogous result for the PSM, where the role of $J^1M $ is played by the cotangent bundle $T^*M$.  Besides the cited literature, gauge symmetries of the model have been studied in \cite{Vancea}.

In \cite{Bascone:2020drt, Bascone:2021njt} two main classes of target spaces have been explicitly considered, namely contact and locally conformal symplectic manifolds. It has been shown that in both cases the auxiliary fields $(\eta,  \lambda)$ can be integrated out and a second-order action description in terms  of the sole embedding maps can be given. Finally, a dynamical extension of the model has been considered, with the introduction of a metric term, similarly to the dynamical extension of the PSM. The auxiliary fields can be integrated out,  yielding a Polyakov action supplemented with a geometric constraint.

Advancements in the field  have led to the exploration of twisted Jacobi structures, which introduce a novel level of complexity to the study of the geometric structures involved as possible string backgrounds. These shall be the object of the present work.

From a mathematical point of view, twisted Jacobi structures can be deduced directly as a generalization of twisted Poisson structures. To this, we briefly review the main steps leading to  the latter \cite{Severa:2001qm}. A Poisson structure on a manifold $M$ naturally defines a   Dirac structure in the standard Courant algebroid $\mathcal{E}=TM \oplus T^*M$.{\footnote{ The vector bundle $\mathcal{E}=TM\oplus T^*M$ is a Courant algebroid (the {\it standard} one), with sections represented by the pairs $(X,\alpha)$, ($X$ and $\alpha$ respectively smooth sections of the tangent and cotangent bundle),  equipped with a nondegenerate inner product  
$$\langle (X,\alpha), (Y,\beta)\rangle = \beta(X)+\alpha(Y),$$  an anchor map $\pi: TM\oplus T^*M\rightarrow TM$ and
a bracket given by
$$[(X,\alpha), (Y,\beta)]= ([X,Y], \mathcal{L}_{X}\beta-\textcolor{red}{\iota_Y}\alpha).$$
The latter is the (non-skewsymmetric) Courant bracket, which satisfies  the Jacobi identity. 
Thus, an $\mathcal{E}$-Dirac structure is a maximally isotropic subbundle of $\mathcal{E}$ with respect to the inner product. }  Indeed, given a bivector field $\Pi$, the graph of $\Pi^{\sharp}: T^*M\rightarrow TM$, with $\Pi^{\sharp}(\alpha)\in \mathfrak{X}(M)$, is a $\mathcal{E}$-Dirac structure  iff $\Pi$ is a Poisson tensor, namely, its Schouten bracket $[\Pi, \Pi]$ vanishes.

However, it is shown in \cite{Severa:2001qm} that one can define a new bracket on $\mathcal{E}$ by using a $3$-form $H$, and the new bracket together with the original bilinear form and anchor on the algebroid $\mathcal{E}$ still constitute a Courant algebroid structure on $TM \oplus T^*M$ if and only if $H$ is closed. This new Courant algebroid is denoted by $\mathcal{E}_{H}$. The bivector $\Pi$ on $M$ previously identified as the Poisson structure associated with $\mathcal{E}$ is now associated with a Dirac structure in the new Courant algebroid $\mathcal{E}_{H}$ if and only if the following relation is satisfied 
\begin{equation}\label{twiPoi}
[\Pi, \Pi]_S= \Pi^{\sharp}H
\end{equation}
where the contraction in $\Pi^{\sharp}H$ is on the first, third and fifth indices of $\Pi \otimes \Pi \otimes \Pi$. 
  \footnote{In local coordinates, we have, up to numerical factors
            $$\Pi^{i r} \partial_r\Pi^{jk}+ cycl(ijk)= \Pi^{pi} \Pi^{qj} \Pi^{sk} H_{pqs}.$$}
The new structure $(\Pi, H)$ is referred to as twisted Poisson structure. In the physics framework it first appeared in the study of topological Poisson sigma models with Wess-Zumino term in \cite{Klimcik:2001vg}, under the name of  WZW-Poisson structure. 

From a mathematical perspective, twisted Jacobi structures were introduced along the same lines, but  using instead the Courant algebroid $\mathcal{E}=(TM \times \mathbb{R})\oplus (T^*M \times \mathbb{R})$ \cite{Nunes:2006a,Cioroianu:2020a}. There,  it is shown that the graph of a given  {bivector field $\Lambda$ on $M$ is a Dirac structure  in the Courant algebroid $(TM \times \mathbb{R})\oplus (T^*M \times \mathbb{R})$ iff $(\Lambda, E)$ is a Jacobi structure, with $E$ the Reeb vector field}. Asking that the fundamental objects of the Jacobi structure be still associated with the Dirac structures of the Courant algebroid with modified brackets, will lead to  modifications of the  Jacobi structures themselves.

In this work we consider a generalization of the previously introduced Jacobi sigma model with the addition of a Wess-Zumino term to accomodate for a twisted Jacobi background as target space. In particular, we consider the five-dimensional sphere $S^5$, that,  as we shall see,  is an interesting example  of twisted contact manifold. As in the non-twisted case, we  will show that it is possible to integrate out   the auxiliary fields for twisted contact manifolds. Finally, we will   present a dynamically-extended model where a metric term is introduced. The analysis is performed both in the general scenario and in the specific example of $S^5$. {Besides being a nontrivial case study, } the latter could have applications to the study of IIB superstrings propagating on the $AdS_{5} \times S^{5}$ background that are dual to planar ${\cal N}=4$ super Yang Mills theory through the celebrated AdS/CFT correspondence. In the context of string theory a  twisted version of JSM is  already considered in \cite{Chatzistavrakidis:2020gpv};  similarly to  the non-twisted case, their model differs from the one introduced here for the presence of an extra scalar field coming from the target space reduction of a twisted PSM. %, eventually shedding light, among other aspects,  on its  integrability.

The paper is organised as follows. In section \ref{sectwistedjacobistructures} we present a brief review of Jacobi and twisted Jacobi structures and describe the particular examples of locally conformal symplectic and contact structures, as well as their twisted counterparts. We also review the construction of $S^5$ as a twisted contact manifold. In section \ref{Sectwistedjacobisigmamodel} the action functional for the twisted Jacobi sigma model is stated. Moreover we present a dynamical extension of the twisted model in which a metric term is introduced, as well as a description of the procedure for the integration of the auxiliary fields,  generalizing the constructions in \cite{Bascone:2020drt, Bascone:2021njt}. In section \ref{Secexamples} we consider the model on the  sphere $S^5$, with twisted structure, in the topological and dynamical versions and  provide the second-order formulation. We conclude with a final discussion of the results with remarks and perspectives.

\section{Twisted Jacobi structures}\label{sectwistedjacobistructures}
In this section we review the main properties of twisted Jacobi structures and twisted Jacobi manifolds, starting from a brief summary on standard Jacobi structures. \\
\subsection{Jacobi structures}\label{Jacobiman}
%\begin{definition}
A Jacobi manifold\footnote{This short review is based on \cite{Vaisman, marle} to which we refer for a detailed analysis.} is a smooth manifold, which is equipped with a Jacobi structure $(\Lambda, E)$, namely  a pair composed by a bivector field $\Lambda \in \Gamma(\largewedge^2 TM)$ and a vector field $E \in \Gamma(TM)$ (the Reeb vector field) satisfying
\begin{equation}\label{usual}
\left[\Lambda, \Lambda \right]_S=2E \wedge \Lambda, \quad \mathcal{L}_E \Lambda=0.
\end{equation}
%\end{definition}
Then,  the Jacobi structure $(\Lambda, E)$ defines a Jacobi bracket on $\mathcal{F}(M)$ that is given by
\begin{equation}\label{jacobibracket}
\{f,g \}_{J}=\Lambda(df,dg)+\iota_E(fdg-gdf), \quad f,g \in \mathcal{F}(M)
\end{equation}
with $\iota_E$ indicating the contraction operator.
The Jacobi bracket  endows the space of smooth functions on $M$, $\mathcal{F}(M)$, with a local Lie algebra structure in the sense of Kirillov;  viceversa, a local Lie algebra on $\mathcal{F}(M)$ induces a Jacobi structure on $M$ \cite{KirillovJacobi}. As can be easily noticed from \eqref{jacobibracket}, Jacobi brackets are linear, skew-symmetric and satisfy Jacobi identity. However, they fail to satisfy Leibniz rule, differently from Poisson brackets. It is clear, however, that Poisson structures can be regarded as a particular case of Jacobi ones when the Reeb vector field is identically vanishing. The Jacobi brackets then reduce to Poisson brackets. 

Jacobi brackets implement the Hamiltonian morphism of Lie algebras $f \mapsto \xi_f$, with 
\begin{equation}\label{hamiltderivationnontwist}
\xi_f=\Lambda^{\sharp} df+fE,
\end{equation}
being
\begin{equation}
\left[\xi_f, \xi_g\right]=\xi_{\{f, g\}_J}, \quad \left[\xi_f, E \right]=\xi_{\mathcal{L}_E f},
\end{equation}
where ${\sharp}: T^*M \to TM$ is the  bundle map induced by the Poisson tensor according to  $\Lambda^{\sharp}(\alpha_{b}\dd x^b)=\Lambda^{ab}\alpha_{b}\del_a$.
 It is extended  to higher forms as 
  $ \Lambda^{\sharp}(\alpha_{a_1\dots a_n}\dd x^{a_1}\wedge\dots\wedge  \dd x^{a_n}) =  \Lambda^{b_1 a_1}\dots \Lambda^{b_n a_n} 
   \alpha_{a_1\dots a_n} \del_{b_1} \wedge \del_{b_n}
  $. 
  
  Eq. \eqn{hamiltderivationnontwist} defines the Hamiltonian vector fields associated with Jacobi brackets.

There are two main classes of Jacobi manifolds, which are represented by locally conformal symplectic manifolds (LCS), which are even-dimensional, and contact manifolds which are odd-dimensional.

\begin{example}
(Locally conformal symplectic structure). Let $M$ be an even-dimensional smooth manifold. The pair $(\Omega, \alpha)$ consisting of a non-degenerate $2$-form $\Omega \in \Omega^2(M)$ and a closed $1$-form $\alpha \in \Omega^1(M)$ (called Lee form) is said to be a locally conformal symplectic manifold (LCS) if
\begin{equation}
d\Omega+\alpha \wedge \Omega=0.
\end{equation}
\end{example}
The unique Jacobi structure $((\Lambda, E), \Omega)$ associated with the locally conformal symplectic structure described above is given by
\begin{equation}\label{lcsformulas}
\iota_E \Omega=-\alpha, \quad \iota_{\Lambda^{\sharp}\gamma} \Omega=-\gamma \quad \forall \gamma \in \Omega^1(M).
\end{equation}
These relations also lead to $\Lambda=\Omega^{-1}$. It is easy to check that symplectic manifolds are a particular case when the Lee form is identically vanishing, moreover, when $\alpha$ is exact, we have globally conformal symplectic manifolds.

\begin{example}
(Contact structure). Let $M$ be an odd-dimensional smooth manifold with $\text{dim}M=2n+1$. The $1$-form $\theta \in \Omega^1(M)$ is said to be a contact structure if $\theta \wedge (d\theta)^n$ is a volume form. The contact form is defined as an equivalence class of $1$-forms up to multiplication by a non-vanishing function. The associated Jacobi structure $(\Lambda, E)$ is given by the following relations
\begin{equation}
\iota_E \theta=1, \quad \iota_Ed\theta=0;
\end{equation}
\begin{equation}
\Lambda^{\sharp}\theta=0, \quad \iota_{\Lambda^{\sharp}\gamma}d\theta=-\gamma+(\iota_E\gamma)\theta \quad \forall \gamma \in \Omega^1(M).
\end{equation}
\end{example}

LCS and contact manifolds are then manifolds equipped with LCS and contact structures respectively.

These two classes of manifolds are particularly relevant as it is possible to show \cite{dazord}   that if a Jacobi structure $(\Lambda, E)$ is transitive, i.e. its characteristic distribution is transitive\footnote{The characteristic distribution of the Jacobi manifold is a distribution whose fibre, at each $x\in M$ is the subspace of $T_x M$ generated by the Hamiltonian vector fields \eqn{hamiltderivationnontwist}. It is said to be transitive if it coincides with the tangent bundle.}, then $(M, (\Lambda, E))$ is either a LCS manifold or a contact manifold.

An important result due to Lichnerowicz \cite{Lich} shows that it is possible to associate with any Jacobi manifold $M$ a higher dimensional Poisson manifold $M \times \mathbb{R}$ with a specific prescription. Since this will hold for twisted structures as well,  we will come back to this point  in Sec. \ref{secjacobiwithbackground}.

\subsection{Twisted Jacobi structures}

\begin{definition}
%Let $M$ be a smooth manifold. 
The pair $((\Lambda, E), \omega)$, with $\Lambda \in \Gamma(\largewedge^2 TM)$, $E \in \Gamma(TM)$  and $\omega \in \Omega^2(M)$ a $2$-form, is said to be a twisted Jacobi structure if the following relations are verified:
\begin{equation}
\begin{aligned}
{} & \frac{1}{2}\left[\Lambda, \Lambda \right]_S-E \wedge \Lambda=\Lambda^{\sharp}d\omega+E \wedge \Lambda^{\sharp}\omega , \\ & \mathcal{L}_E \Lambda= \left[\Lambda, E \right]_S=-\left(\Lambda^{\sharp} \iota_E d\omega+\Lambda^{\sharp} \iota_E \omega \wedge E \right).
\end{aligned}
\end{equation}
\end{definition}
In the case $\omega=0$ one recovers the usual relations for a Jacobi structure \eqn{usual}. % and then $M$ is a Jacobi manifold.

A twisted Jacobi structure endows the  space  of functions $\mathcal{F}(M)$ with a $\mathbb{R}$-linear and skew-symmetric bracket $\{\cdot , \cdot\}_{tJ}: \mathcal{F}(M) \times \mathcal{F}(M) \to \mathcal{F}(M)$ which is formally the same as \eqn{jacobibracket}
\begin{equation}
\{f,g \}_{tJ}=\Lambda(df,dg)+\iota_E(fdg-gdf), \quad f,g \in \mathcal{F}(M)
\end{equation}
the difference with \eqn{jacobibracket} being that it has now a   non-vanishing Jacobiatior (i.e. twisted Jacobi brackets no longer satisfy Jacobi identity):
\begin{equation}\label{twistedjacobiidentity}
\begin{aligned}
\operatorname{Jac}\{f, g, h\} & =\iota_{\Lambda^{\sharp} d\omega+E \wedge \Lambda^{\sharp} \omega }(df \wedge dg \wedge dh) \\
& -\iota_{\Lambda^{\sharp} \iota_E d\omega+\Lambda^{\sharp} \iota_E \omega \wedge E}(fdg \wedge dh+gdh \wedge df+hdf \wedge dg),
\end{aligned}
\end{equation}
with $f,g,h \in \mathcal{F}(M)$.
Since the twisted Jacobi structure is still a first-order differential operator in each entry, we define the Hamiltonian vector fields  as in  Eq. \eqref{hamiltderivationnontwist}.% for  the twisted case.
%, namely $\xi_f=\Lambda^{\sharp}df+fE$. 
Therefore  one can rewrite \eqref{twistedjacobiidentity} more explicitly as
%\begin{equation}\label{twistedjacobiidentityexpl}
%\begin{aligned}
%\operatorname{Jac}\{f,g,h\}=
%& -\iota_{\xi_f \wedge \xi_g \wedge \xi_h} d\omega+\left(\mathcal{L}_E f\right) \iota_{\xi_g \wedge \xi_h} \omega \\
%    & +\left(\mathcal{L}_E g\right) \iota_{\xi_h \wedge \xi_f} \omega+\left(\mathcal{L}_E h\right)  \iota_{\xi_f \wedge  \xi_g} \omega.
%\end{aligned}
%\end{equation}
\begin{equation}\label{twistedjacobiidentityexpl}
%\begin{aligned}
\operatorname{Jac}\{f,g,h\}=
 -\iota_{\xi_f \wedge \xi_g \wedge \xi_h} d\omega+\left(\mathcal{L}_E f\right) \iota_{\xi_g \wedge \xi_h} \omega 
   +\left(\mathcal{L}_E g\right) \iota_{\xi_h \wedge \xi_f} \omega+\left(\mathcal{L}_E h\right)  \iota_{\xi_f \wedge  \xi_g} \omega.
%\end{aligned}
\end{equation}
However, differently from before,  the map $f \mapsto \xi_f$ is not a homomorphism of Lie algebras. We have indeed
\begin{equation}\label{liehom1}
%\begin{aligned}
%
{\left[\xi_f, \xi_g\right]-\xi_{\{f, g\}} } %& 
=\Lambda^{\sharp} \iota_{\xi_f \wedge \xi_g} d\omega-\left(\mathcal{L}_E f\right) \Lambda^{\sharp} \iota_{\xi_g} \omega 
%\\& 
+\left(\mathcal{L}_E g\right) \Lambda^{\sharp} \iota_{\xi_f} \omega+\left(\iota_{\xi_f \wedge \xi_g} \omega\right) E
%\end{aligned}
\end{equation}
and
\begin{equation}\label{liehom2}
\left[\xi_f, E\right]+\xi_{\mathcal{L}_E f}=\Lambda^{\sharp}\left(\iota_{\xi_f \wedge E} d\omega-\left(\mathcal{L}_E f\right) \iota_E \omega\right)+\left(\iota_{\xi_f \wedge E} \omega\right) E.
\end{equation}
Analogously to the non-twisted case, we can define the characteristic distribution generated by the Hamiltonian vector fields
\begin{equation}
\mathcal{C}_{((\Lambda, E), \omega)}(M)=\bigcup_{x \in M}\left\langle\left\{\left(\xi_f\right)_x: f \in \mathcal{F}(M)\right\}\right\rangle \subseteq T M,
\end{equation}
that is involutive because of \eqref{liehom1} and \eqref{liehom2} \footnote{A distribution is called involutive if its sections form a Lie subalgebra.}. 
Hence,  a twisted Jacobi manifold is said to be transitive if its characteristic distribution is transitive, i.e. it coincides, at each point, with the tangent space.

\begin{example}
(Twisted locally conformal symplectic structure). Let $M$ be an even-dimensional smooth manifold. The pair $(\Omega, \alpha)$ consisting of a non-degenerate $2$-form $\Omega \in \Omega^2(M)$ and a closed $1$-form $\alpha \in \Omega^1(M)$ is said to be a locally conformal symplectic manifold twisted by the $2$-form $\omega \in \Omega^2(M)$ if 
\begin{equation}
d(\Omega-\omega)+\alpha \wedge (\Omega-\omega)=0.
\end{equation}
The unique twisted Jacobi structure $((\Lambda, E), \omega)$ associated with the twisted locally conformal symplectic structure described is given by the same expressions as in \eqref{lcsformulas}, i.e. 
\begin{equation}
\iota_E \Omega=-\alpha, \quad \iota_{\Lambda^{\sharp}\gamma} \Omega=-\gamma \quad \forall \gamma \in \Omega^1(M)\end{equation}
and it naturally leads to $\Lambda=\Omega^{-1}$.
\end{example}

\begin{example}
(Twisted contact structure). Let $M$ be an odd-dimensional smooth manifold with $\text{dim}M=2n+1$. The contact structure $\theta \in \Omega^1(M)$ is said to be twisted by the $2$-form $\omega \in \Omega^2(M)$ if $\theta \wedge (d\theta+\omega)^n$ is a volume form. The associated twisted Jacobi structure $((\Lambda, E),\omega)$ is given by the following conditions:
\begin{equation}\label{twistedcontactcond1}
\iota_E \theta=1, \quad \iota_E(d\theta+\omega)=0;
\end{equation}
\begin{equation}\label{twistedcontactnocoord}
\Lambda^{\sharp}\theta=0, \quad \iota_{\Lambda^{\sharp}\gamma}(d\theta+\omega)=-\gamma+(\iota_E\gamma)\theta \quad \forall \gamma \in \Omega^1(M),
\end{equation}
which fix uniquely the bivector $\Lambda$ and the twisted Reeb vector field $E$. It will also be useful for applications to write explicitly in coordinates the relations in \eqref{twistedcontactnocoord}:
\begin{equation}\label{twistedcontactcoord}
\Lambda^{ij}\theta_j=0, \quad \Lambda^{ij}(d\theta+\omega)_{jk}=-\delta^{ik}+E^i \theta_k.
\end{equation}
\end{example}

These two examples of twisted Jacobi manifolds can be regarded as having the same role and importance of the LCS and contact manifolds for the non-twisted case, since we have the two following results {\cite{Nunes:2006b}}
\begin{theorem}
Let $M$ be a smooth manifold. If a twisted Jacobi structure $((\Lambda, E),\omega)$ is transitive, i.e. its characteristic distribution is transitive, then $M$ is either a twisted locally conformal symplectic manifold or a twisted contact manifold.
\end{theorem}
\begin{theorem}
The characteristic distribution of a twisted Jacobi structure is completely integrable, with the characteristic leaves being either twisted locally conformal symplectic manifolds or twisted contact manifolds.
\end{theorem}

\subsection{Jacobi structures with background}\label{secjacobiwithbackground}

\begin{definition}
A pair $((\Lambda, E),(H, \omega))$, with $\Lambda \in \Gamma(\largewedge^2 TM)$, $E \in \Gamma(TM)$, $H \in \Omega^3(M)$ and $\omega \in \Omega^2(M)$ satisfying the relations
\begin{equation}\label{jacobibackgroundrel}
\begin{aligned}
{} & \frac{1}{2}\left[\Lambda, \Lambda \right]_S-E \wedge \Lambda=\Lambda^{\sharp}H+E\wedge\Lambda^{\sharp}\omega , \\ & \mathcal{L}_E \Lambda= \left[\Lambda, E \right]_S=-\left(\Lambda^{\sharp} \iota_E H+\Lambda^{\sharp} \iota_E \omega \wedge E \right),
\end{aligned}
\end{equation}
is called a Jacobi structure  with background $(H,\omega)$, or a relaxed twisted Jacobi structure {\cite{Cioroianu:2020a}}.
\end{definition}
It is immediate to verify that, if we take the $3$-form $H$ to be exact and $H=d\omega$, then we recover twisted Jacobi structures. This means that any twisted Jacobi structure $((\Lambda, E), \omega)$ can be regarded as a Jacobi structure $(\Lambda, E)$ with background $(d\omega, \omega)$. 

Note also that, as one would expect, a twisted Poisson structure $(\Pi, H)$, verifying Eq. \eqn{twiPoi}, can be considered as a Jacobi structure with background $((\Pi, E=0),(H,\omega))$.

Finally, let us notice that for  Jacobi structures with background, Eqs.  \eqref{twistedjacobiidentity}-%\eqref{twistedjacobiidentityexpl}, \eqref{liehom1}, 
\eqref{liehom2} remain valid on  replacing $d\omega$ with $H$.

\begin{example}
(Locally conformal symplectic structure with background). Let $M$ be an even-dimensional smooth manifold. The locally conformal symplectic structure $(\Omega, \alpha)$ is said to be a locally conformal symplectic manifold with background $(H,\omega)$ if 
\begin{equation}\label{relaxedlcs}
H=d\Omega+\alpha \wedge (\Omega-\omega).
\end{equation}
The unique associated Jacobi structure $(\Lambda, E)$ with background $(H,\omega)$ is given by
\begin{equation}
\iota_E \Omega=-\alpha, \quad \iota_{\Lambda^{\sharp}\gamma} \Omega=-\gamma \quad \forall \gamma \in \Omega^1(M).
\end{equation}
It is obvious from \eqref{relaxedlcs} that the locally conformal symplectic structure with background $((\Omega,\alpha),(d\omega,\omega))$ is just the twisted one $((\Omega, \alpha),\omega)$.
\end{example}

Note that twisted contact structures $(\theta,\omega)$ can be regarded as contact structures with background $(\theta,(d\omega,\omega))$.

Maintaining the transitivity definition, the characteristic distribution corresponding to a Jacobi structure with background enjoys analogous properties to the ones for the standard twisted cases. In particular, we have {\cite{Cioroianu:2020a}}:

\begin{theorem}
Let $M$ be a smooth manifold. If a Jacobi structure with background $((\Lambda, E),(H,\omega))$ on $M$ is transitive, then $M$ is either a locally conformal symplectic manifold with background or a twisted contact manifold.
\end{theorem}
\begin{theorem}
The characteristic distribution of a Jacobi structure with background is completely integrable, with the characteristic leaves being either locally conformal symplectic manifolds with background or twisted contact manifolds.
\end{theorem}

It might be surprising that there is no contact structure with {non-trivial} background. However, this lack of structure is clarified as a result of the following gauge theorem \cite{Cioroianu:2020a}

\begin{theorem}
Let $((\Lambda, E),(H_1,\omega_1))$ and $((\Lambda, E),(H_2, \omega_2))$ be two transitive Jacobi structures with background on the smooth manifold $M$. Then, the following alternative cases hold:
\begin{enumerate}
\item $M$ is even-dimensional: there exists a $2$-form $\omega \in \Omega^2(M)$ such that \footnote{$\Lambda^{\flat}$ denotes the map $\Lambda^{\flat}: TM \to T^*M$ such that $\Lambda^{\sharp}\Lambda^{\flat}=\mathds{1}_{TM}$.}
\begin{equation}
\omega_1=\omega_2+\omega, \quad H_1=H_2-\omega \wedge \Lambda^{\flat}E;
\end{equation}
\item $M$ is odd-dimensional: 
\begin{equation}
\omega_1=\omega_2, \quad H_1=H_2.
\end{equation}
\end{enumerate}
\end{theorem}
The result of this theorem provides the gauge transformations for a transitive Jacobi manifold with background, i.e. the changes of $H$ and $\omega$ that do not modify the Jacobi structure $(\Lambda, E)$. In this sense the fact that contact manifolds with background do not exist is elucidated by the fact that for a twisted contact structure $(\theta, \omega)$ which is transitive and odd-dimensional, the background $(d\omega, \omega)$ cannot be changed. 

%\begin{example}
%(Locally conformal symplectic structures with background).
%\end{example}

It is important to remark that a Poissonization procedure also works for twisted Jacobi structures and Jacobi structures with background. More explicitly, we have the following {\cite{Cioroianu:2020a}}
\begin{theorem}
Given a Jacobi structure with background $((\Lambda, E),(H,\omega))$ on the smooth manifold $M$, the product manifold $\tilde{M}=M \times \mathbb{R}$ carries a Poisson structure with background $(P, \phi)$ defined by 
\begin{equation}
P=e^{-\tau}\left(\Lambda+\frac{\partial}{\partial \tau} \wedge E \right), \quad \phi=e^{\tau}\left(H+d\tau\wedge \omega \right), \quad \tau \in \mathbb{R}.
\end{equation}
\end{theorem}
and
\begin{corollary}
Given a twisted Jacobi structure $((\Lambda, E),\omega)$ on the smooth manifold $M$, the product manifold $\tilde{M}=M \times \mathbb{R}$ carries a twisted (exact) homogeneous Poisson structure $(P, \phi)$ defined by 
\begin{equation}
P=e^{-\tau}\left(\Lambda+\frac{\partial}{\partial \tau} \wedge E \right), \quad \phi=d(e^{\tau} \omega), \quad \tau \in \mathbb{R}.
\end{equation}
\end{corollary}
The corollary follows from the fact that for $H=d\omega$ then $e^{\tau}\left(H+d\tau\wedge \omega \right)=d(e^{\tau} \omega)$.

Analogously to the symplectification in the case of Poissonization of a contact structure, here we have that if $(\theta, \omega)$ is a twisted Jacobi structure on $M$, then the Poissonization procedure leads to a twisted symplectification. In fact, on $M \times \mathbb{R}$ we would have non-degenerate $P$ with inverse $\Omega=e^{\tau}(d\theta+\omega+d\tau \wedge \omega)$ and $\phi=d(e^{\tau}\omega)=d\Omega$. So we have a twisted symplectic structure on $M \times \mathbb{R}$. Of course, in the case $\omega=0$ one recovers the symplectification of a contact structure.

\subsection{The sphere $S^5$ as a twisted contact manifold}\label{S5}

It is not easy to find examples of twisted contact structures. But it is easy to check that  almost cosymplectic (ACoS) manifolds are actually  twisted contacts manifold and viceversa.

Cosymplectic structures are relevant both for mathematical \cite{Cappelletti-Montano:2013hka}  and for physical applications where they play a major role in  geometric formulations of time-dependent mechanics \cite{deLeon:2020hnm,Atesli:2022ira,bercen:2023}. 

\begin{definition}\cite{Cappelletti-Montano:2013hka}
An almost cosymplectic structure on a $(2n+1)$-dimensional manifold $M$ is a pair $(\theta, \Omega) \in \Omega^1(M) \times \Omega^2(M)$ such that $\mu=\theta \wedge \Omega^n$ is a volume form. The structure is said to be  cosymplectic  if $\theta$ and $\Omega$ are both closed. 
For an almost cosymplectic manifold there exists a vector field $E$ such that $\iota_E \Omega=0$ and $\iota_E \theta=1$. 
\end{definition}
Note that in the literature sometimes ACoS structures are also called almost contact structures.

It can be easily  noticed that a twisted contact structure $(\theta, \omega)$, as defined in Sec. \ref{sectwistedjacobistructures}, can be considered as an almost cosymplectic structure with $\Omega=d\theta+\omega$. Viceversa, an almost cosymplectic structure $(\theta, \Omega)$ can be viewed as a twisted contact structure with $\omega=\Omega-d\theta$. Instances of ACoS manifolds can be found e.g. in \cite{Libermann:1962}, where a few examples like the sphere $S^5$ and some quadrics are worked out.

%We will be interested in $S^5$ in particular, for physical reasons, so, following \cite{Libermann:1962}, we will review its construction as an ACoS manifold, and then as a twisted contact manifold.

 Following \cite{Libermann:1962}, one starts with  a manifold $N_{2n+2}$ equipped with an almost symplectic structure $\gamma \in \Omega^2(N_{2n+2})$\footnote{ We recall that an almost symplectic structure on an even-dimensional manifold    is a two-form $\gamma$ that is everywhere non-singular. If in addition $ \gamma$  is closed then it defines a  symplectic structure.} and a regular submanifold $M_{2n+1}$ defined by $f=k$, with $k$ a constant and $f \in C^{\infty}(M_{2n+2})$. The vector field $V$ tangent to $M_{2n+2}$ such that $\iota_V \gamma=df$ has no singularity on $M_{2n+1}$ and the restriction $W=V |_{M_{2n+1}}$ is tangent to $M_{2n+1}$. The restriction $\Omega= \gamma|_{M_{2n+1}}$ is a form of rank $2n$ and satisfies $\iota_{W}\Omega=0$.  It is   the pullback of  $\gamma$ through the embedding $i \, : \, M_{2n+1}  \xhookrightarrow{} M_{2n+2} $. Given a Riemannian metric $g$ on $M_{2n+1}$, let $E$ be a norm-one vector field (w.r.t. $g$) collinear to $W$. 
Then, the 1-form $\theta$ is uniquely defined by the conditions $\iota_E \theta = 1$, $i_E \Omega=0$ and $\theta \wedge \Omega^n$  being a volume form.
 %If $\theta$ is the $1$-form dual to the vector field $E$, then: $\iota_E \theta=1$, $\iota_E \Omega=0$ and $\theta \wedge \Omega^n \neq 0$ everywhere, i.e. it is a volume form. 
 These relations endow $(M_{2n+1},\theta, \Omega)$ with an  almost cosymplectic structure.

Let us consider now the sphere $S^5$ with the following realisation as a submanifold \cite{Libermann:1962},:
\begin{equation}
S^5 \underset{\scriptstyle{X^1=0}}{\subset} S^6 \underset{\sum_i X^i X^i=1}{\subset} \mathbb{R}^7, \quad \{X^i \}_{i\in \{1,\dots,7 \}} \in \mathbb{R}^7.
\end{equation}
The almost complex (hence almost symplectic) structure on $S^6$ induces an ACoS structure on $S^5$ via the procedure described above\footnote{We need the embedding in $S^6$ and not just in $\R^6$, in order to get a non-degenerate two-form on the sphere $S^5$.}. In particular, one finds
\begin{equation}
E(X)=X^4 \frac{\partial}{\partial X_2}+X^7\frac{\partial}{\partial X_3}-X^2\frac{\partial}{\partial X_4}+X^6\frac{\partial}{\partial X_5}-X^5\frac{\partial}{\partial X_6}-X^3\frac{\partial}{\partial X_7}
\end{equation}
and
\begin{equation}
\begin{aligned}
\Omega(X){} & =X^2(dX^5 \wedge dX^3+dX^7 \wedge dX^6)+X^4(dX^7 \wedge dX^5+dX^3 \wedge dX^6) \\ & +X^3(dX^6 \wedge dX^4+dX^2 \wedge dX^5) +X^7(dX^6 \wedge dX^2+dX^5 \wedge dX^4) \\ & +X^5(dX^4 \wedge dX^7+dX^3 \wedge dX^2)+X^6(dX^2 \wedge dX^7+dX^4 \wedge dX^3).
\end{aligned}
\end{equation}
The $1$-form $\theta$ can then be obtained by imposing the condition $\iota_E \theta=1$ (as long as $\theta \wedge \Omega^n$ is a volume form):
\begin{equation}
\theta(X)=X^4dX^2+X^7dX^3-X^2dX^4+X^6dX^5-X^5dX^6-X^3dX^7,
\end{equation}
so that the pair $(\theta, \Omega)$ form an ACoS structure.

It is useful to relabel the coordinates as
\begin{equation}
\begin{aligned}
{} & X^2 \to y^1, \quad X^3 \to y^2, \quad X^4 \to x^1 \\ & 
X^5 \to y^3, \quad X^6 \to x^3, \quad X^7 \to x^2,
\end{aligned}
\end{equation}
so to obtain the compact expressions:
\begin{eqnarray}
\theta&=&\sum_{i=1}^3 \left(x^i dy^i-y^idx^i\right) \\
E&=&\sum_{i=1}^3 \left(x^i \frac{\partial}{\partial y^i}-y^i \frac{\partial}{\partial x^i}\right) \\
\Omega&=& \sum_{\underset{\text{mod}(3)}{i=1}}^3 {}  \Big[x^i(dx^{i+1} \wedge dy^{i+2}+dy^{i+1} \wedge dx^{i+2})\nonumber\\
&+& y^i(dx^{i+1}\wedge dx^{i+2}+dy^{i+2}\wedge dy^{i+1}) \Big].
\end{eqnarray}
On introducing complex combinations $z^i=x^i+iy^i$ and the complex conjugate $\bar z^i=x^i-iy^i$, with $i=1,2,3$ the latter become \begin{equation}\label{thetas5}
\theta=\frac{i}{2}\sum_{i=1}^3 \left(z^i d\bar{z}^i-\bar{z}^i dz^i \right),
\end{equation}
\begin{equation}\label{reebs5}
E=i \sum_{i=1}^3 \left(z^i \partial_i-\bar{z}^i \bar{\partial}_{i} \right),
\end{equation}
\begin{equation}
\Omega=\frac{i}{4} \epsilon_{ijk}\left(z^i dz^j \wedge dz^k-\bar{z}^i d\bar{z}^j \wedge d\bar{z}^k \right),
\end{equation}
where we introduced the holomorphic and antiholomorphic derivatives respectively as $\partial_i=\frac{1}{2}\left(\frac{\partial}{\partial x^i}-i \frac{\partial}{\partial y^i} \right)$, $\bar{\partial}_i=\frac{1}{2}\left(\frac{\partial}{\partial x^i}+i \frac{\partial}{\partial y^i} \right)$.

Because of the one-to-one relation between ACoS and twisted manifolds, we have that $(S^5, (\theta, \omega))$ is a twisted contact manifold with 
\begin{equation}\label{omegas5}
\omega=\Omega-d\theta=\frac{i}{4} \left( \epsilon_{ijk}z^i dz^j \wedge dz^k-\epsilon_{ijk}\bar{z}^i d\bar{z}^j \wedge d\bar{z}^k -4\delta_{jk} dz^j \wedge d\bar{z}^k\right).
\end{equation}
In order to read the Jacobi structure $((\Lambda, E), \omega)$ of $S^5$ from its twisted contact structure $(\theta, \omega)$, we have to determine the bivector $\Lambda$ by using the relations in \eqref{twistedcontactcoord}. The result is given by
\begin{equation}\label{lambdas5}
\Lambda=-2i {\epsilon_i}^{jk}\left(\bar{z}^i \partial_j \wedge \partial_k - z^i \bar{\partial}_j \wedge \bar{\partial}_k \right).
\end{equation}
whereas the Reeb vector field and the twisting two-form $\omega$ are respectively given by Eqs. \eqn{reebs5} and \eqn{omegas5}.

In the next section we introduce the Jacobi sigma model for a generic twisted Jacobi manifold, whereas in Sec. \ref{Secexamples} we will finally consider the model on $S^5$.

\section{Twisted Jacobi sigma model}\label{Sectwistedjacobisigmamodel}
The purpose of this section is to generalise the construction of the twisted Poisson sigma model \cite{Klimcik:2001vg} to twisted Jacobi structures. 
We recall that the former was obtained by adding to the Poisson sigma model action a twisting term, thus yielding
\begin{equation}\label{twistedpoisson}
S_P[X, A]=\int_{\Sigma=\partial V} \left[A_i \wedge dX^i+\frac{1}{2}\Pi^{ij}(X)A_i \wedge A_j \right]+\int_{V} \tilde{X}^*H,
\end{equation}
with $X: \Sigma \rightarrow M$ the base map and $A \in \Omega^1\left(\Sigma, X^*\left(T^*M \right)\right)$. The field $\tilde{X}$ represents an extension of the field $X$ on $V$, such that $\tilde{X}|_{\Sigma}=X$. Here the twisting term is defined on a $3$-manifold $V$ whose boundary is $\Sigma$. The bivector $\Pi$ and the 3-form  $H$ satisfy Eq. \eqn{twiPoi}, namely, they define a twisted Poisson structure on $M$.

By following the same principles in mind for the Poisson sigma model, in order to twist the Jacobi sigma model we modify the action represented  by Eq. \eqn{jacobac} with the addition of a Wess Zumino term,  in complete analogy with the Poisson case. We therefore define
\begin{equation}\label{twistedjacobiaction}
S[X, \eta, \lambda]=\int_{\Sigma=\partial V} \left[\eta_i \wedge dX^i+\frac{1}{2}\Lambda^{ij}(X)\eta_i \wedge \eta_j-E^i(X) \eta_i \wedge \lambda\right]+\int_{V} \tilde{X}^*H.
\end{equation}
%with field configurations represented by $(X, \Xi)$,  $X: \Sigma \rightarrow M$ the base map and $\Xi=(\eta, \lambda) \in \Omega^1\left(\Sigma, X^*\left(T^*M \oplus \mathbb{R}\right)\right)$. 
As in   \eqn{twistedpoisson}, the field $\tilde{X}$ is an extension of the field $X$ on $V$, with  $\tilde{X}|_{\Sigma}=X$, being $V$ again a $3$-manifold whose boundary is $\Sigma$ . The background fields $((\Lambda, E), (H,\omega))$ form a Jacobi structure with background and satisfy the relations in \eqref{jacobibackgroundrel}. Note that the background $3$-form $H$ has  to be closed in order to  obtain the desired  two-dimensional contribution to the equations of motion from the Wess-Zumino term. An action similar to \eqn{twistedjacobiaction} was already considered in \cite{Chatzistavrakidis:2020gpv}, although with and extra scalar field, already present in their version fo the JSM, which is related to the homogeneous Poisson structure  on $M\times \R$ where a PSM can be built.

The equations of motion, obtained by variations of the action, read 
\begin{equation}\label{eom1}
dX^i+\Lambda^{ij}\eta_j-E^i \lambda=0
\end{equation}
\begin{equation}\label{eom2}
d\eta_i +\frac{1}{2}\partial_i \Lambda^{jk}\eta_j \wedge \eta_k-\partial_i E^j \eta_j \wedge \lambda+\frac{1}{2}H_{ijk} dX^j \wedge dX^k=0
\end{equation}
\begin{equation}\label{eom3}
E^i \eta_i=0.
\end{equation}
By comparison with the non-twisted case, one observes that   the only equation   which gets modified  is the second one. 
When imposing the consistency between the three equations, namely applying the exterior derivative to Eq. \eqref{eom1} and implementing the other two, as well as the defining conditions on the background fields, another equation for $\lambda$ is obtained:
\begin{equation}\label{eom4}
d\lambda=\frac{1}{2}\Lambda^{ij}\eta_i \wedge \eta_j +\frac{1}{2}\omega_{ij}\,dX^i \wedge dX^j.
\end{equation}
This differs from the non-twisted one for the presence of the  $2$-form $\omega$.

\subsection{Integration on twisted contact manifolds}

In what follows we  integrate out the auxiliary fields $\eta$ and $\lambda$ in the special case of twisted contact manifolds,  obtaining a second order action for the sole configuration fields  $X$. We have seen in Sec. \ref{secjacobiwithbackground} that  there is no generic pair $(H, \omega)$  which can twist the contact structure,  but it has to be specifically $H=d\omega$. In this case, the action \eqref{twistedjacobiaction} reduces to 
\begin{equation}\label{twistedexactjacobiaction}
S[X, \eta, \lambda]=\int_{\Sigma} \left[\eta_i \wedge dX^i+\frac{1}{2}\Lambda^{ij}(X)\eta_i \wedge \eta_j-E^i(X) \eta_i \wedge \lambda+\omega_{ij} \, dX^i \wedge dX^j \right].
\end{equation}
The equations of motion  are given by  \eqref{eom1}-\eqref{eom4}  with the replacement $H_{ijk}=\partial_i \omega_{jk}$.
These can be solved for $\eta$ and $\lambda$, thanks to Eqs.  \eqref{twistedcontactcond1}, \eqref{twistedcontactcoord},  satisfied by the twisted contact structure. In particular, let us contract Eq. \eqref{eom1} with $\theta_i$ so to have
\begin{equation}
\theta_i(dX^i+\Lambda^{ij}\eta_j-E^i \lambda)=0.
\end{equation}
Because of \eqref{twistedcontactcond1} we are left with
\begin{equation}
\lambda=\theta_idX^i=\theta.
\end{equation}
In order to solve for $\eta$ we can contract  the same equation,  now with $(d\theta+\omega)_{ij}$, yielding 
\begin{equation}
(d\theta+\omega)_{ij}dX^j+(d\theta+\omega)_{ij}\Lambda^{jk}\eta_k=0,
\end{equation}
where use has been made of  the second equation in \eqref{twistedcontactcond1}. By using the second equation in \eqref{twistedcontactcoord} and the constraint \eqref{eom3} we are finally left with
\begin{equation}
\eta_i=(d\theta+\omega)_{ij}dX^j.
\end{equation}
Now we can put the expressions for $\eta$ and $\lambda$ back into the action \eqref{twistedexactjacobiaction}. By using the properties of   the twisted contact structure, after some manipulations we obtain  the following second-order action:
\begin{equation}\label{twistedcontactintegratedaction}
S_2=-\frac{1}{2} \int_{\Sigma} X^*(d\theta-\omega)=\frac{1}{2}\int_{\Sigma} X^* B.
\end{equation}
Where we have introduced the B-field notation $B=\omega-d\theta$. Note that the $2$-form $B$ is closed only if $\omega$ is closed, namely $H=0$. If $\omega$ is not closed,  $H=d\omega=dB$, represents  the H-flux of $B$.

\subsection{Twisted dynamical extension to the Jacobi sigma model}

In \cite{Bascone:2020drt, Bascone:2021njt} we showed that, on  introducing  a dynamical extension of  the topological action, the integration of  the auxiliary fields yields a Polyakov action supplemented by a geometric constraint with background metric and B-field, $(g, B)$, which depend on the Jacobi structure. Here we follow the same procedure, by modifying  the action according to
\begin{equation}
S=\int_{\Sigma=\partial V}\left[\eta_i \wedge d X^i+\frac{1}{2} \Lambda^{i j}(X) \eta_i \wedge \eta_j-E^i(X) \eta_i \wedge \lambda+\frac{1}{2}\left(G^{-1}\right)^{i j}(X) \eta_i \wedge \star \eta_j\right]+\int_V \tilde{X}^* H,
\end{equation}
where $G$ is the metric tensor on the target (twisted Jacobi) manifold $M$. The Hodge star operator $\star$ implements the metric on the worldsheet $\Sigma$ which we take to be $h=\text{diag}(1,-1)$.
The new equations of motion are
\begin{equation}\label{dynamicaleom1}
d X^i+\Lambda^{i j} \eta_j-E^i \lambda+\left(G^{-1}\right)^{i j} \star \eta_j=0,
\end{equation}
\begin{equation}\label{dynamicaleom2}
d \eta_i+\frac{1}{2} \partial_i \Lambda^{j k} \eta_j \wedge \eta_k-\partial_i E^j \eta_j \wedge \lambda+\frac{1}{2}H_{ijk}\, dX^j \wedge dX^k+\frac{1}{2} \partial_i\left(G^{-1}\right)^{j k} \eta_j \wedge \star \eta_k=0,
\end{equation}
\begin{equation}
E^i \eta_i=0
\end{equation}
The only difference with the non-twisted dynamical case is the presence of the $H$ term in Eq. \eqref{dynamicaleom2}. However, the latter does not affect the integration process of the auxiliary fields performed in \cite{Bascone:2020drt, Bascone:2021njt}. Moreover, the Wess-Zumino term present in the action  is already written in terms of the sole embedding maps. Therefore, we repeat   the  same procedure as  in \cite{Bascone:2020drt, Bascone:2021njt}.

Being $G$ naturally non-degenerate we can solve \eqn{dynamicaleom1}
 for  $\star \eta$, 
\begin{equation}\label{eqstareta}
\star \eta_j=-G_{ij}\left(dX^i+\Lambda^{ik}\eta_k-E^i \lambda \right)
\end{equation}
and obtain $\eta$ by applying   the Hodge star to the latter 
\begin{equation}\label{eqeta}
\eta_p=-{(M^{-1})^j}_p G_{ij}\left(\star dX^i-\Lambda^{ik}G_{\ell k} dX^{\ell}+\Lambda^{ik}G_{\ell k}E^{\ell}\lambda-E^i \star \lambda \right),
\end{equation}
with  
\be\label{matrixm}
{M^p}_j={\delta^p}_j-G_{j i}\Lambda^{ik}G_{k \ell}\Lambda^{\ell p}
\ee
  a   symmetric matrix, which we take to be non-degenerate irrespective of the  rank  of  $\Lambda$ by assumption. 
The action becomes then 
\begin{equation}\label{actionlambda}
\begin{aligned}
S(X, \lambda)& = \int_{\Sigma=\partial V} {}  \bigg[\frac{1}{2}{(M^{-1})^p}_i G_{jp} \, dX^i \wedge \star dX^j- \frac{1}{2}{(M^{-1})^p}_i G_{\ell p} \Lambda^{\ell k} G_{jk} \, dX^i \wedge dX^j \\ & - \frac{1}{2}{(M^{-1})^p}_i G_{\ell p}\Lambda^{\ell k} G_{mk} E^m \lambda \wedge dX^i+ \frac{1}{2}{(M^{-1})^p}_i G_{\ell p} E^{\ell} \star \lambda \wedge dX^i \bigg]+\int_V \tilde{X}^* H  .
\end{aligned}
\end{equation}
The remaining auxiliary field, $\lambda$, is integrated out by   
using  the inner product of one-forms, 
\begin{equation}
\int_{\Sigma} \star \lambda \wedge dX=-\int_{\Sigma} \lambda \wedge \star dX,
\end{equation}
so that  in \eqn{actionlambda} $\lambda$ becomes  a Lagrange multiplier imposing the geometric constraint
\begin{equation}\label{polyakovconstraint}
(M^{-1})_{i \ell}\left( \Lambda^{\ell k}G_{mk}E^m dX^i+E^{\ell} \star dX^i\right)=0.
\end{equation}
Therefore,  the term proportional to  $\lambda$ vanishes on-shell and we are left with the second order action
\begin{equation}
S=\int_{\Sigma=\partial V} \left(g_{ij}\,dX^i \wedge \star dX^j+B_{ij}\, dX^i \wedge dX^j \right)+\int_V \tilde{X}^* H,
\end{equation}
supplemented with the geometric constraint \eqn{polyakovconstraint}, 
%\begin{equation} \label{polyakovconstraint}
%\left(M^{-1}\right)_{i \ell}\left(\Lambda^{\ell k} G_{m k} E^m d X^i+E^{\ell} \star d X^i\right)=0,
%\end{equation}
}
with 
\begin{equation}\label{gB}
g_{i j}=G_{j p}\left(M^{-1}\right)^p{ }_i \, , \quad B_{i j}=G_{i k}\left(M^{-1}\right)^p{ }_j G_{p \ell} \Lambda^{\ell k}.
\end{equation}
%where we defined 
%\begin{equation}\label{matrixm}
%M^p{ }_j=\delta^p{ }_j-G_{j i} \Lambda^{i k} G_{k \ell} \Lambda^{\ell p}.
%\end{equation}
%It can be shown that $M$ is a symmetric matrix, which we assume to be non-degenerate irrespective of the rank of $\Lambda$.
When $H$ is exact, as in  the case of twisted contact manifolds where  $H=d\omega$,  the Polyakov action becomes
\begin{equation}
S=\int_{\Sigma} \left(g_{ij}\,dX^i \wedge \star dX^j+B_{ij}\, dX^i \wedge dX^j \right),
\end{equation}
where the metric $g$ is the same as in \eqref{gB}, but the $B$-field is now given by
\begin{equation}\label{gbclosedh}
B_{i j}=G_{i k}\left(M^{-1}\right)^p{ }_j G_{p \ell} \Lambda^{\ell k}+\omega_{ij}.
\end{equation}

\section{The twisted Jacobi sigma model on $S^5$}\label{Secexamples}
The procedure described in the previous section may be applied to $(S^5, \theta, \omega)$, that has been shown to be a twisted contact manifold in Sec. \ref{S5}.  

 By using the Jacobi structure $((\Lambda, E),\omega)$ of the twisted $5$-sphere given by Eqs.  \eqref{reebs5}, \eqref{omegas5}, \eqref{lambdas5} ,   the action \eqref{twistedexactjacobiaction} becomes
\begin{equation}\label{actionons5}
\begin{aligned}
S=\int_{\Sigma} { } &  \zeta_i dz^i +\bar{\zeta}_i d\bar{z}^i+i{\epsilon_i}^{jk}\left(z^i \bar{\zeta}_j \wedge \bar{\zeta}_k-\bar{z}^i \zeta_j \wedge \zeta_k \right)+i\left(\bar{z}^i \bar{\zeta}_i-z^i \zeta_i \right) \wedge \lambda \\ & +\frac{i}{4}\epsilon_{ijk}\Big(z^i dz^j \wedge dz^k-\bar{z}^i d\bar{z}^j \wedge d\bar{z}^k \Big)-i\delta_{j\bar{k}}dz^j \wedge d\bar{z}^{\bar{k}}+\Phi \left(\delta_{i\bar{j}} z^i \bar{z}^{\bar{j}} -1\right)
\end{aligned}
\end{equation}
Note that we have explicitly added   the constraint $\sum_{i=1}^3 z^i \bar{z}^i=1$, with $\Phi$ a Lagrange multiplier,  implementing the embedding in $\C^3$. %because we are parametrizing the sphere $S^5$ using $\mathbb{R}^6$ coordinates. %This is necessary in this case since $S^5$ is not parallelizable. 
The field $\Phi$ will be then a $2$-form on $\Sigma$ and a scalar on the target. The Greek symbols  $\zeta$ represent the momenta in the new complex basis, and they are complex linear combinations of the original $\eta$, as it is showed in the following relations:
\begin{equation}
\begin{aligned}
{} & \zeta_1=\frac{1}{2}(\eta_4-i\eta_2) , \quad \bar{\zeta}_1=\frac{1}{2}(\eta_4+i\eta_2) \\ &
\zeta_2= \frac{1}{2}(\eta_7-i\eta_3), \quad \bar{\zeta}_2=\frac{1}{2}(\eta_7+i\eta_3) \\ &
\zeta_3= \frac{1}{2}(\eta_6-i\eta_5), \quad \bar{\zeta}_3=\frac{1}{2}(\eta_6+i\eta_5).
\end{aligned}
\end{equation}
The remaining field $\lambda$ is instead real. This can also be easily understood in terms of the Poissonized action with target space the product manifold $S^5 \times \mathbb{R}$ . In fact, the latter may be parametrized in terms of $((z,\bar{z}), u)$, with $(z,\bar{z}) \in \C^3$ satisfying $\sum_{i=1}^3 z^i\bar{z}^i=1$ and $u \in \mathbb{R}$. Therefore, when considering the Poisson sigma model with target  $S^5 \times \mathbb{R}$, the field   $\eta_u$ is  the conjugated variable to $u \in\mathbb{R}$. When restricting the model to   $S^5$ as described in \cite{Bascone:2020drt, Bascone:2021njt},  $\lambda$ appears related to the  original $\eta_u$ via the pull-back:  $\eta_u=\pi^* \lambda$, with $\pi: S^5\times \R\rightarrow S^5$ the projection. Notice that $\lambda$   is a scalar on the target albeit a $1$-form on the source.
 
\subsection{Second-order topological and Polyakov actions on twisted $S^5$}

By using the result in \eqref{twistedcontactintegratedaction}, we can write the second-order topological action for the twisted $S^5$ explicitly. In fact, by using the expressions for $\theta$ and $\omega$ in \eqref{thetas5} and \eqref{omegas5} respectively, we can write
\begin{equation}\label{topactions5}
S_2=-\frac{1}{2}\int_{\Sigma} \Big[2i \delta_{ij} dz^i \wedge d\bar{z}^j-\frac{i}{4} \epsilon_{ijk} z^i dz^j \wedge dz^k+\frac{i}{4}\epsilon_{ijk}\bar{z}^i d\bar{z}^j \wedge d\bar{z}^k+\Phi\left(\delta_{ij} z^i \bar{z}^j-1 \right) \Big].
\end{equation}
Note that we had to introduce again the field $\Phi \in \Omega^2(\Sigma)$ as a Lagrange multiplier imposing the constraint $\delta_{ij}z^i \bar{z}^j=1$.

For the dynamical model on $S^5$ with the parametrization $(z,\bar{z}) \in \mathbb{C}^3$ with $\sum z \bar{z}=1$, we can introduce the metric tensor on $S^5$ as 
\begin{equation}
G_{i \bar{j}}=\frac{1}{2} \delta_{i \bar{j}},
\end{equation}
given by $ds^2=\delta_{i\bar{j}}dz^i \otimes d\bar{z}^j$ with $\delta_{i\bar{j}}z^i \bar{z}^j=1$, so that, together with symmetry properties of the structures involved we will have a Polyakov action of the general form
\beqa
S&=&\int_{\Sigma} \left(g_{ij} \,dz^i \wedge \star dz^j+ g_{i \bar{j}} \,dz^i \wedge \star d\bar{z}^j+B_{ij}\,dz^i \wedge dz^j + B_{i \bar{j}} \,dz^i \wedge d\bar{z}^j +c.c. \right)\nonumber\\
&+&\int_{\Sigma} \Phi(\delta_{ij}z^i \bar{z}^j-1)+ \mathcal{C}
\eeqa
where $\mathcal{C}$ is the geometric constraint \eqn{polyakovconstraint}.
The matrix $M$ from  \eqref{matrixm} results to be 
\begin{equation}
M_{i \bar{j}}= \delta_{i \bar{j}}-\frac{1}{2}z^m \bar{z}^{\bar{n}}\delta_{i \bar{n}} \delta_{m \bar{j}}.
\end{equation}
Notice that the latter has only mixed indices since $G$ has only mixed indices while $\Lambda$  has only  indices of the same type (see \eqref{lambdas5}). With this result, the geometric constraint reads 
\begin{equation}
\mathcal{C}=\lambda \operatorname{Tr}\left(\delta_{i \bar{j}} \, \bar{z}^{\bar{j}} \star dz^i \right), 
\end{equation}
which is trivially satisfied when implementing  $\sum_{i=1}^3 z^i \bar{z}^i=1$.
Finally, by using Eq. \eqref{gbclosedh} the background $(g,B)$ reads
\begin{equation}
\begin{aligned}
{} & g_{ij}=0 ,\quad g_{i \bar{j}}=\frac{1}{4}\left(\delta_{i\bar{j}} +z^p \bar{z}^{\bar{q}} \delta_{i\bar{q}}\delta_{p\bar{j}}\right), \\ & 
 B_{ij}=\frac{i}{8} \epsilon_{ijk}z^k, \quad B_{i\bar{j}}=\omega_{i\bar{j}}=-i\delta_{i\bar{j}}.
\end{aligned}
\end{equation}
Notice that the topological part of the action is exactly the action in \eqref{topactions5}. Notice also that the metric in complex coordinates has the form of a Euclidean version of a Kerr-Schild metric, which is a deformation of Minkowski metric usually written as $g_{ab}=\eta_{a b}+\phi(x)k_{a}k_{b}$, where $\eta$ is the Minkowski metric and $k_a$ are null vectors (w.r.t. both $\eta$ and $g$), with $\phi$ any scalar function of the coordinates.

\section{Conclusions and Outlook}
\label{sectconclusions}

We have considered a twisted version of the Jacobi sigma model with the addition of a Wess-Zumino term, to take into account twisted Jacobi manifolds and Jacobi manifolds with background as target spaces, the latter ones being generalizations of the twisted Poisson backgrounds. The twisted Jacobi sigma model is then a two-dimensional topological non-linear gauge field theory describing strings moving on twisted Jacobi backgrounds. 

In particular, we have considered twisted contact backgrounds as target spaces and shown that in these cases the auxiliary fields can be always integrated out and a second-order description in terms of the sole embedding maps can be given. The resulting action is a topological sigma model where the $B$-field is given by a combination of the exterior derivative of the contact form and the twisting $2$-form. The $B$-field  is closed only if the twisting $2$-form is closed. The sphere $S^5$ has been considered as an explicit example of twisted contact manifold, the twisted contact structure being derived from its almost cosymplectic structure. 

Finally, we have introduced a dynamical extension of the model, which is obtained by adding a metric term to the action functional. On integrating out the auxiliary fields, a Polyakov action is obtained, with a metric and B-field, supplemented by a geometric constraint, all of them  explicitly derived  in terms of the twisted Jacobi structure. As an explicit example we considered again the twisted contact sphere $S^5$, with a resulting Polyakov action which contains a metric that in complex coordinates has the form of a Euclidean version of a Kerr-Schild metric.

As a future  direction of research it would be interesting to study the quantization of the model, which could be approached within geometric quantization, taking advantage of existing results \cite{deleon}, or, analogously to the (twisted) Poisson sigma  model, by exploring deformation quantization of Jacobi structures, which, due to the twisting should involve non-associativity. 

Finally, it would be interesting to investigate the possibility to formulate the model on a line bundle, similarly to what has been done for the Jacobi sigma model
in \cite{dc}.

 \vspace{5pt}

%\noindent{\bf Acknowledgements} 
%
%...

%\begin{appendix}%\label{appA}
% \appendixpage
%\addappheadtotoc
%\section{Appendix: }\label{appA}
%\section{Appendix}

%%%%%%%%%%%%%%%%%%%%%%%%%%%%%%%%%%%%%%%%%%%%%%
%%%%%%%%%%%%%%%%%%%%%%%%%%%%%%%%%%%%%%%%%%%%%
%%%%%%%%%%%%%%%%%%%%%%%%%%%%%%%%%%%%%%%%%%%%%%%

%\subsection
%\appendix

%\end{appendix}

\end{document}